\documentclass[aps,pra,twocolumn,showpacs,psfig]{revtex4}

\usepackage{graphicx}
\usepackage{amsmath}
\usepackage{amssymb}
\usepackage{epsfig}
\begin{document}
\title {Dynamics of matter solitons in weakly modulated optical lattices}
\author{V.A. Brazhnyi$^1$} 
\email{brazhnyi@cii.fc.ul.pt}
\author{V.V. Konotop$^{1,2}$} \email{konotop@cii.fc.ul.pt}
\author{V. Kuzmiak$^3$} \email{kuzmiak@ure.cas.cz}

\affiliation{$^1$
Centro de F\'{\i}sica Te\'{o}rica e Computacional, Universidade de Lisboa, 
Complexo Interdisciplinar, Av. Prof. Gama Pinto 2, Lisbon 1649-003, Portugal
\\
$^2$ Departamento de F\'{\i}sica,
Universidade de Lisboa, Campo Grande, Ed. C8, Piso 6, Lisboa
1749-016, Portugal
\\
$^3$Institute of Radio Engineering and Electronics, Czech
Academy of Sciences, Chaberska 57, 182 51 Prague 8, Czech Republic}
    

\begin{abstract}
It is shown that matter solitons can be effectively managed by means of smooth variations of parameters of optical lattices in which the condensate is loaded. The phenomenon is based on the effect of lattice modulations  on the carrier wave transporting the soliton and that is why is well understood in terms of the effective mass approach, where a particular spatial configuration of the band structure is of primary importance. Linear, parabolic, and spatially localized modulations are considered as the case examples. It is shown that these defects can originate accelerating and oscillating motion of matter solitons as well as simulate soliton interaction with attractive and repulsive defects.
\end{abstract}

\maketitle

\section{Introduction}

Realization of a Bose-Einstein condensate (BEC) in an optical lattice~\cite{AndKas} originated intensive experimental and theoretical studies~\cite{group,gapsol,instab,LZ,LZ1,AndKas,Morsch,lens,Eiermann,Scott,CEL,GapSol,AKS,YSR} 
of the phenomenon. One of the  
main features induced by the periodicity is the appearance of a band structure in the 
spectrum of the underlying linear system, i.e. the gas of noninteracting atoms. 
The band spectrum is responsible for a number of effects, including modulation instability\cite{gapsol,instab} and formation of gap solitons 
\cite{gapsol}, Landau-Zener tunneling \cite{LZ,LZ1}, Bloch oscillations of BEC's \cite{LZ,AndKas,Morsch}, lensing effect~\cite{lens}, soliton stabilization \cite{Eiermann}, etc. These effects, being based on the properties of the linear system, are well described in terms of the concept of the effective mass, which takes into account wave nature of the phenomenon and that is why its inverse value is also referred to as the group velocity dispersion~\cite{gapsol}. In practice, laser lattices are never perfect. In particular, they are often imposed simultaneously with other external potentials, like, for example, a magnetic trap which significantly affects the motion of a soliton~\cite{Scott}. Moreover one can create spatially localized lattices, propagation of matter waves through which displays a number of interesting properties~\cite{CEL} 
such as resonant transmission and soliton generation through the modulational instability. It is also highly relevant to mention that very recently direct experimental observation of a gap matter soliton in $^{87}$Rb BEC, containing about 250 atoms, has been reported~\cite{GapSol}.

In this paper we present analysis of the BEC dynamics in an optical lattice with smoothly modulated parameters. In this case the band gap structure is still preserved but is deformed by the modulation. In particular, we show that this is a way of managing the dynamics of matter waves - making them accelerating, oscillating, etc. 

The model and the physics of the phenomenon are described in Sec.\ref{model}. Acceleration of a matter soliton in a linearly modulated lattice is considered in Sec.~\ref{accel}. Soliton oscillations in a lattice with a parabolic modulation of the depth are considered in Sec.~\ref{sec:parab}. Sec.~\ref{sec_defect} is devoted to soliton interaction with a spatially localized defects. The results of the paper are summarized in the Conclusion.

\section{Physics of the phenomenon} 
\label{model}
 
Let us consider a trap potential which can be written down 
in the form $V_{trap}=\frac m2(\omega_\|^2x^2+\omega_0^2y^2+\omega_0^2z^2)+V_\epsilon(x)$.   The first term in the right hand side of this expression 
describes a magnetic trap with $\omega_\|$ and $\omega_0$ being the longitudinal and transverse 
linear oscillator frequencies. The condensate is chosen to have a cigar shape with the aspect ratio 
satisfying the relation $\omega_\|/\omega_0\ll a_0^2/\xi^2\ll 1$ ($a_0$ and $\xi$ being the transverse 
linear oscillator length and the healing length, respectively). We are interested in excitations of a BEC 
having characteristic scales of order of the healing length and a relatively small amplitude, what 
allows us to neglect the term $\frac m2\omega_\|^2x^2$ in the expression for the trap potential $V_{trap}$.  The term $V_\epsilon(x)$, where $\epsilon$ 
is a deformation parameter controlling the potential shape, describes smoothly modulated optical lattice. It
will be assumed to have the form $V_\epsilon(x)\equiv f(\epsilon^{3/2} x)V_0(x)$, where $V_0(x)=V_0(x+L)$ 
is an unperturbed lattice having a period $L$ of order of the linear oscillator length $a_0$, $L\sim a_0$, and 
$f(\epsilon^{3/2} x)$ is a smooth modulation normalized as follows: $f(0)=1$. Smoothness in the present context 
means slow variation of $f(\epsilon^{3/2}x)$ on the scale of the healing length, or in other words $\epsilon\sim a_0/\xi$. 
Then, one can apply the multiple-scale expansion~\cite{gapsol} in order to reduce the Gross-Pitaevskii equation governing the dynamics of the BEC to an effectively 
one-dimensional (1D) nonlinear Schr\"{o}dinger (NLS) equation with the slowly varying effective 
mass, $ m_\alpha\equiv m_\alpha(\epsilon^{1/2}X)=[\partial_k^2 
{\cal E}_\alpha(k;\epsilon^{1/2}X)]^{-1}$, 
and the group velocity of the carrier wave depending in the coordinate, $v_\alpha=\partial_k 
{\cal E}_\alpha(k;\epsilon^{1/2}X)$:
\begin{eqnarray}
	\label{GP1}
	i\Psi_T+iv_\alpha\Psi_X= 	-(2m_\alpha)^{-1}\Psi_{XX} 	+\chi |\Psi|^2\Psi.
\end{eqnarray}
Hereafter $\alpha$ and $k$ stand for a number of the zone and for the dimensionless wave vector 
in the first Brillouin zone (BZ), respectively, $\chi=$sign$\,a_s$, $a_s$ being the s-wave scattering length. 
In (\ref{GP1}) we have 
passed to dimensionless slow variables: $X= \epsilon x/a_0=\epsilon\tilde{x}$ and 
$T=\epsilon^2 \omega_0t=\epsilon^2 \tilde{t}$.  Taking into account smoothness of the potential, the 
spectrum, ${\cal E}_\alpha(k;\epsilon^{3/2}\tilde{x})$ in some point of the space, say $\tilde{x}=\tilde{x}_0$ ($\tilde{x}_0= x_0/a_0$), can  be computed from the linear eigenvalue problem 
\begin{equation}
\label{Hill}
\frac{d^2 \varphi_{\alpha k}}{d\tilde{x}^2} + \left[{\cal E}_\alpha(k;\epsilon^{3/2}\tilde{x}_0) - 
f(\epsilon^{3/2} \tilde{x}_0)V_0(\tilde{x})\right]\varphi_{\alpha k} =0.
\end{equation}
 
In order to get qualitative picture of effects which can be observed in the matter wave dynamics we carry out numerical 
simulations using a 1D model (justified for for the low-density BEC's)
\begin{equation}
\label{nls}
i\psi_{\tilde{t}}= - \psi_{\tilde{x}\tilde{x}} + f(\epsilon^{3/2}\tilde{x})V_0(\tilde{x})\psi 
+\chi |\psi|^2\psi.
\end{equation}
It is to be emphasized here that $\psi$ in Eq.~(\ref{nls}) describes the complete macroscopic wave function whereas $\Psi$ in Eq.~(\ref{GP1})  
describes behavior of its envelope.

In the case of a homogeneous lattice, i.e.  Eq. (\ref{nls}) with $f(\tilde{x})\equiv 1$, there exists a solitary 
wave solution (bright matter soliton) which can be found by means of the stationary ansatz  
$\psi(\tilde{x},\tilde{t})=u(\tilde{x})e^{-i{\cal E} \tilde{t}}$, where $u(\tilde{x})$ is a real 
valued function satisfying the equation as follows
\begin{equation}
\label{Mathieu0}
u_{\tilde{x}\tilde{x}} + \left[{\cal E} - V_0(\tilde{x})\right]u -\chi u^3=0.
\end{equation}
 
Let us consider now the boundary of the BZ, i.e. $|k_0|=\pi a_0/L$. As it is known (see e.g. \cite{AKS}) 
stationary solitary wave solutions subject to zero boundary conditions 
$\lim_{|\tilde{x}|\to \infty} u(\tilde{x},\tilde{t}) = 0$, exist only if ${\cal E}$ belongs to a gap 
of the spectrum of (\ref{Hill}). For the sake of definiteness, below we are dealing with the first lowest forbidden gap, which lower and upper edges will be denoted 
as ${\cal E}^{(1)}$  and ${\cal E}^{(2)}$,  respectively (see an example in Fig.~\ref{fig1}d). 
Thus the frequency of a static matter soliton must satisfy the condition ${\cal E}^{(1)}<{\cal E}< {\cal E}^{(2)}$. 
Then  in the case $\chi=-1$ ($\chi=1$) a small amplitude bright matter soliton can be excited with energies inside the gap in the vicinity of the upper, ${\cal E}^{(2)}$ (lower, ${\cal E}^{(1)}$) band edges, correspondingly~\cite{gapsol}. 
For the next consideration it is important to mention that  envelope solitons can also exist at 
${\cal E}>{\cal E}^{(2)}$ if $\chi=-1$, and  at ${\cal E}<{\cal E}^{(1)}$ if $\chi=1$. Respective solitons are created against a moving carrier wave background (with $|k|\neq |k_0|$ and $v_\alpha\neq 0$), i.e. they 
move with the velocity $v_\alpha$, and are described by the formula 
\begin{eqnarray}
\label{solit}
\Psi_s(X,T)=\frac{\sqrt{-\chi m_\alpha}N\exp\left(\frac{i}{8}m_\alpha N^2T \right)}
{2\cosh\left[\frac 12 m_\alpha N\left(X-X_0(T) \right)\right]},
\end{eqnarray}
with either $\chi=-1$, $\alpha=2$, and  $m_{2}>0$, or  $\chi=1$, $\alpha=1$, and  $m_{1}<0$. 
$X_0(T)=-v_\alpha T$ is a coordinate of the soliton center and $N$ is a normalized number of atoms
\begin{eqnarray}
\label{N}
N=\int_{-\infty}^{\infty}|\Psi|^2dX.
\end{eqnarray}

Strictly speaking Eq.~(\ref{solit}) is a solution of the unperturbed NLS equation (\ref{GP1}) with $v_\alpha$ 
and $m_\alpha$ constants. Since, however, the parameters are changing slowly in space, the simplified picture 
based on the analytic form (\ref{solit}) appears to be good enough for qualitative understanding various phenomena.

In what follows the consideration is restricted to the case $\chi=-1$ and discuss the dynamics of small-amplitude bright matter soliton near the upper edge of the first band gap ${\cal E}^{(2)}$.


\section{Matter wave acceleration.}
\label{accel}

\begin{figure}[h]
\includegraphics[width=8.5cm,clip]{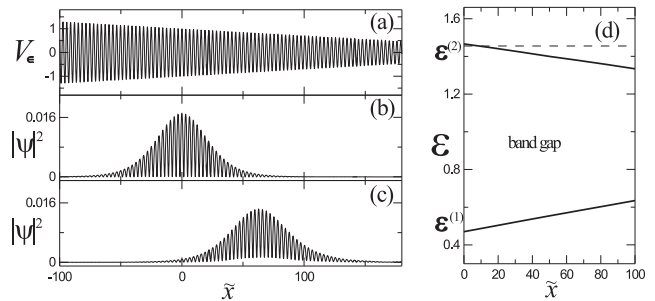}
\caption{(a) The linearly modulated periodic potential $V_{\epsilon}$ given by (\ref{lin}).   
(b) The initial profile of the condensate near the upper bound of the gap for ${\cal E}^{(2)}=1.46$. (c) The profile of the condensate at time $t=100$. (d) The band 
structure the potential (\ref{lin}) (solid line) and the energy of the soliton (dashed line). All data are computed for  $\epsilon=0.02$.}
\label{fig1}
\end{figure}

 We first consider linear modulation of the lattice amplitude (see Fig.~\ref{fig1}a)
\begin{equation}
\label{lin}
V_\epsilon = (1-\epsilon^{3/2} \tilde{x}) \cos(2\tilde{x}).
\end{equation}
which lowest forbidden gap is shown on Fig.~\ref{fig1}d.
\begin{figure}[h]
\includegraphics[width=8.5cm,clip]{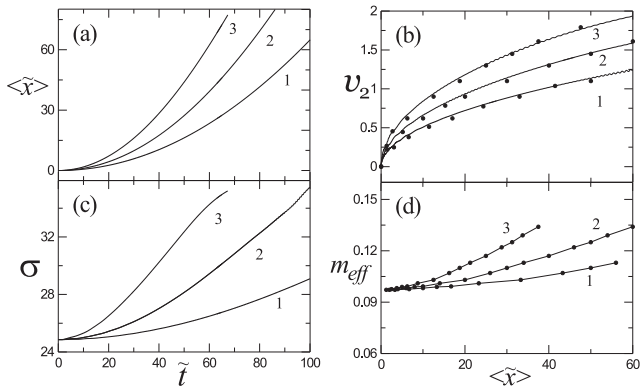}
\caption{(a) The coordinate of the soliton center of mass,  
$\left\langle \tilde{x}\right\rangle$, {\it vs} time.
(b) Dependences of the carrier wave group velocity $v_2$ obtained directly from the periodic structure 
(dots) and velocity of the soliton center of mass obtained from the direct numerical simulations 
(solid line).
(c) Evolution of the dispersion of the soliton. 
(d) Dependence of the effective mass $m_{2}$ on the soliton coordinate. Curves 
1, 2 and 3 correspond to $\epsilon=0.02, 0.03$, and $0.04$.}
\label{fig2}
\end{figure}

Let us assume that in the vicinity of the origin, i.e. near 
$\tilde{x}=0$, a bright static gap soliton with $v_2=0$ is created (see Fig.~\ref{fig1}b). Due to nonzero extension of the soliton, it partially occupies 
the space where the soliton energy falls into allowed band  (see Fig.~\ref{fig1}d)  and 
thus corresponds to running linear waves. This is a reason for a soliton to start to move toward 
the region with  a narrower gap,  i.e. in the positive direction.  
Since it is assumed that in the leading approximation the energy of the soliton is conserved,  
the group velocity of the background  increases and, therefore,  soliton velocity grows when 
the coordinate of the soliton center increases (Fig.~\ref{fig2}a), what gives rise to a  soliton acceleration 
(see Fig.~\ref{fig2}b). An example of dynamics of the coordinate of the condensate center of mass, $\left\langle \tilde{x}\right\rangle$, and its  dispersion $\sigma=\left(\left\langle 
\tilde{x}^2\right\rangle-\left\langle \tilde{x}\right\rangle^2\right)^{1/2}$, obtained by numerical integration of Eq.~(\ref{nls}), are shown in 
Fig.~\ref{fig2} (a) and (c) (angular brackets 
stand for spatial average). From this figure one can see that the soliton indeed undergoes 
acceleration in such a way that the soliton velocity follows the change of the group velocity 
of the background. 
Simultaneously, the dispersion of the soliton increases as it is shown in Fig.~\ref{fig2}c. 
From the first sight  this result does not resemble a model given by Eq.~(\ref{solit}) 
because according to this model the soliton width should decrease as the effective mass 
increases.

\begin{figure}[h]
\includegraphics[width=8.5cm,clip]{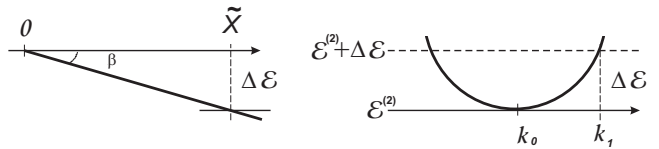}
\caption{Schematic illustration of the calculus of the  dependence of the group velocity and the effective mass on the 
coordinate, for linearly modulated potential. The notations are defined in the text.}
\label{fig_add}
\end{figure}

In order to explain the apparent discrepancy let us consider   the dispersion relation of the second zone, ${\cal E}_2(k)$, in the vicinity of the band edge in more detail. Since ${\cal E}_2(k)$  is an even function of the wave 
vector, one can expand  
\begin{eqnarray}
\label{expansion}
{\cal E}_2(k)={\cal E}^{(2)} +\frac 12 \partial_k^2 {\cal E}_2(k_0)\cdot (k-k_0)^2
\nonumber \\
+\frac{1}{4!} \partial_k^4 {\cal E}_2(k_0)\cdot (k-k_0)^4,
\end{eqnarray}
where ${\cal E}^{(2)}={\cal E}_2(k_0)$.
We assume that  (i) the center of soliton is displaced from the point $\tilde{x} = 0$, 
where its energy was ${\cal E}_2$, to some point $\tilde{x} =\tilde{X}$ without change of the energy, and that
ii) the gap is large enough and within the range $0 < \tilde{x} < \tilde{X}$ in the leading order one    can neglect the change of 
the functional dependence of ${\cal E}_2$ on the wave vector $k$.
Then one has the situation illustrated in Fig.~\ref{fig_add}. When soliton moves from $\tilde{x}=0$ to $\tilde{x}=\tilde{X}$, the acquired energy shift, $\Delta {\cal E}$, toward the allowed zone can be estimated as  $\beta \tilde{X}$, where $\beta$ is a 
small ($\beta\ll 1$) angle between the slope of the band-gap edge and the $x$-axis corresponding to the 
energy ${\cal E}_2$ (see the left figure in Fig.~\ref{fig_add}). The same energy shift on the  
diagram ${\cal E}_2(k)$, giving the energy ${\cal E}^{(2)}+\Delta{\cal E}$, corresponds to the wave vector $k_1$, which 
can be approximately found from the equation $(k_1-k_0)^2 =2m_2\Delta{\cal E}$: 
$k_1=k_0+\sqrt{2m_2\beta \tilde{X}}$. Hence, one computes that dependence of the group velocity on a coordinate $\tilde{x}$ 
is given by 
\begin{eqnarray}
\label{velocity}
v_2=\left\{
\begin{array}{ll} 
\sqrt{\frac{2\beta \tilde{x}}{m_2}} & \tilde{x}>0
\\
0, & \tilde{x}\leq 0
\end{array}\right.
\end{eqnarray}
while the dependence of the effective mass is approximated by 
\begin{eqnarray}
\label{mass}
m_2(\tilde{x})=m_{20}- m_{20}^2\tilde{x}^2\Delta m,\quad \Delta m=\frac{\beta}{6} m_{20}\partial_k^4 {\cal E}_2(k_0)
\end{eqnarray} 
where $m_{20}$ is the effective mass at the boundary of the zone, i.e. at $\tilde{x}=0$
(notice that it follows from Fig.~\ref{fig2}d $\partial_k^4 {\cal E}_2(k_0)<0$). 

It is remarkable that the above simplified estimates give rather precise estimates for the relevant quantities. Indeed, for $\epsilon=0.02$ 
from Fig.~\ref{fig1}d and Fig.~\ref{fig2}d one obtains $\beta\approx 0.0013$ and $m_2\approx 0.1$ what 
gives for $\tilde{X}=30$ an estimate $v_2(\tilde{X}=30)\approx 0.88$, in accordance with (\ref{velocity}) while the numerical 
value is $v_2(\tilde{X}=30)\approx 0.87$ (see  Fig.~\ref{fig2}b).

Returning to the soliton dynamics shown in Figs.~\ref{fig1}, \ref{fig2}, let us neglect small change 
of the effective mass, i.e. make a  substitution $m_\alpha=m_{\alpha 0}$ in the equation for slowly varying amplitude (\ref{GP1}), and compute the 
following integral identities
\begin{eqnarray}
	\label{Nt}
	&&
	\frac{dN}{dT}=\int_{-\infty}^{\infty}(v_\alpha)_X|\Psi|^2dX;
	\\
	&&\frac{dP}{dT}=0,\qquad P=-i\int_{-\infty}^{\infty}\overline{\Psi}\Psi_X dX;
\\
	\label{Pt}
 	&& \frac{d\langle X^2\rangle }{dT}=\frac{i}{m}\int_{-\infty}^{\infty}X(\overline{\Psi}_X\Psi-\overline{\Psi}\Psi_X)dX
	\nonumber \\
	&& +\int_{-\infty}^{\infty}(X^2v_\alpha)_X|\Psi|^2dX, 
      	 \nonumber
      		\,\,\,\,
      		\langle X^2\rangle =\int_{-\infty}^{\infty}X^2|\Psi|^2dX  
      		\\
      		\label{Xt}
\end{eqnarray}
where $N$ is the number of particles given by (\ref{N}), $P$ is the wave momentum in the frame moving with the 
group velocity of the carrier wave, and $\sqrt{\langle X^2\rangle/N}$ is an average width of the wave packet. Then it follows from (\ref{Pt}) that the total momentum 
is preserved. Moreover for a soliton ansatz (\ref{solit}) it is zero, what reflects the fact that the condensate is moving with the group velocity of the carrier wave what corroborates with the findings shown in Fig.~\ref{fig2}b. 
From (\ref{Nt}) it follows that the number of particles is not preserved. 
This, however does not contradict the conservation law of the 
total number of atoms in the Gross-Pitaevskii equation since the model given by the Eq. (\ref{GP1}) describes 
the main approximation of the ground state and does not 
account for the high-frequency radiation while, on the other hand, in the leading order, the soliton is moving in the inhomogeneous medium and thus represents radiating matter wave~\cite{YSR}. Considering the second band, 
$\alpha=2$, and approximating the solution by (\ref{solit}), (\ref{velocity}) one can rewrite (\ref{Nt}) as
\[
\frac{dN}{dT}=\sqrt{\frac{\beta}{2m_{20}}}\int_{0}^{\infty}\frac{dX}{\sqrt{X}}
\left(|\Psi_s(X,T)|^2-|\Psi_s(0,T)|^2\right) \leq 0
\]
what reflects the fact that the soliton looses particles. Finally, taking into account that $\Psi_s$, given 
by (\ref{solit}) is real, one obtains from (\ref{Xt})~
\[
\frac{d\langle X^2\rangle}{dt}=5\sqrt{\frac{\beta}{2m_{20}}}\int_{0}^{\infty}X^{3/2} |\Psi_s(X,T)|^2 dX>0
\] 
what means increase of the soliton width, observed in Figs.~\ref{fig1}c and \ref{fig2}c.
 
The wave packet during its motion 
follows the carrier wave and thus rapidly acquires  
relatively large group velocity which is of order of unity.
The latter factor does not allow adiabatic adjustment of the soliton parameters to increase of the effective mass of the carrier wave (see (\ref{mass}) and Fig.~\ref{fig1}d). Nonadiabaticity of the process manifests itself in radiative losses and broadening of the wave packet.
This  process continues as soliton moves to the region with small potential depth, as it is observed in Fig.~\ref{fig1}c, and with time, as matter of fact, the wave looses its solitonic properties transforming itself in a dispersive wave packet. It worth pointing out that another simple way to look at the broadening of the wave packet is to take into account that soliton wave front has larger velocity than the velocity of its tail leading to spreading of the pulse.    

\section{Oscillation of the matter wave in a lattice subject to parabolic modulation.}
\label{sec:parab}

\begin{figure}[h]
\includegraphics[width=8.5cm,clip]{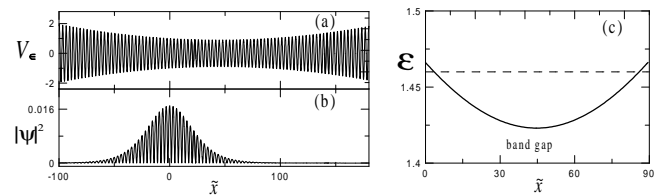}
\caption{ (a) The inhomogeneous periodic potential $V_\epsilon$ given by (\ref{parab}) with $\epsilon=0.02$ and $\Delta \tilde{x}=45$. (b) The initial profile  of an envelope soliton (the same as in Fig.~\ref{fig1}a). (c) The upper edge of the first band gap for for the potential shown in (a). 
}
\label{prof_parab}
\end{figure}

Let us consider now a soliton dynamics (the initial profile is shown in Fig.~\ref{prof_parab}b) in 
a lattice modulated by a parabolic function (see Fig.~\ref{prof_parab}a)
\begin{equation}
\label{parab}
V_\epsilon (\tilde{x})= \left[1+\epsilon^{5/2} (\tilde{x}-\Delta \tilde{x})^2\right] \cos(2\tilde{x})
\end{equation}
where $\Delta \tilde{x}$ is introduced for an initial relative shift between the soliton center and the minimum of the parabolic modulation. The upper bound of the first gap of this potential for $\epsilon=0.02$ is depicted in 
Fig.~\ref{prof_parab}c. 

\begin{figure}[h]
\includegraphics[width=6cm,clip]{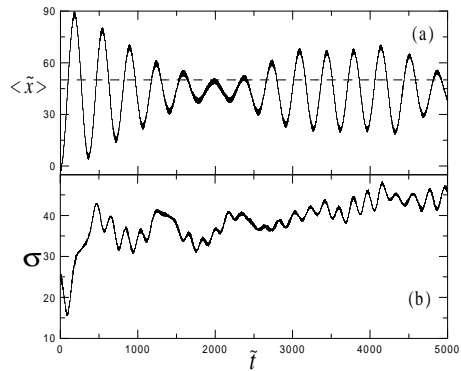}
\caption{  Dynamics of the center of mass (a) and of dispersion (b) of the matter soliton. 
Parameters the same as in Fig.~\ref{prof_parab}. The dashed line corresponds to  
$\langle \tilde{x}\rangle=\Delta \tilde{x}$.}
\label{oscill}
\end{figure}

Now one observes  oscillations of the condensate cloud   (see Fig.~\ref{oscill}). 
The accelerating part of this motion when the soliton is moving toward the center of the potential is explained in the previous Section. The soliton having passed the central part of the potential is decelerating and at some point the velocity of the center of mass of the cloud becomes zero. This occurs in the vicinity 
of the turning point, $x_{turn}$, where the energy of the soliton falls into the forbidden 
gap (the intersections of dashed and solid lines in Fig.~\ref{oscill}c). Indeed, considering a soliton moving from the right to the left, it follows from Eq.~(\ref{velocity}),
that the velocity of the background, and hence the velocity of the soliton, goes to zero as 
$\sqrt{x-x_{turn}}$, when the coordinate of the soliton center approaches  the turning point $x_{turn}$.
Thus one can speak about Bragg reflection of the soliton, from  inhomogeneous periodic potential.
 
There are two features of the soliton dynamics to point out in this case. First, in contrast to 
the case of the linear modulation, at the initial stages the dispersion decreases (see Fig~\ref{oscill}b). 
This compression of the pulse is explained
by its large extension, when the leading part of the wave 
packet undergoes Bragg reflection  and moves in the negative direction while the soliton center still 
moves in the positive direction. The second feature demonstrated
in Fig.~\ref{oscill}a can be identified as  beatings of the mean coordinate of the wave packet. 
In order to explain this effect, we assume that the solution can be described 
approximately by the formula (\ref{solit}) where the coordinate of the soliton 
center $X_0(T)$ depends periodically on time. This dependence is 
determined by the effective linear oscillator frequency $\nu$ associated with the effective 
parabolic potential, the latter being estimated as $  \nu= 2\epsilon^{5/4}$($\approx 0.014$ in our simulations). 
This frequency approximately coincides with the higher frequency that can be 
seen in Fig.~\ref{oscill}a, and which is $\sim 0.017$. The observed discrepancy is 
due to the fact that one deals with a modulated periodic potential, rather than with a 
real parabolic one. On the other hand, the envelope soliton is characterized by the 
internal frequency  $\omega_{int}=m_2N^2/8$ [see Eq.~(\ref{solit})]. In our numerical 
simulations it is a low frequency $\omega_{int}\approx 0.002$. Thus by expanding the modulus 
of the soliton solution in the Fourier series with respect to time the motion is 
determined by superposition of the lowest harmonics with $\nu\pm\omega_{int}$ which results in 
the beating phenomenon. The above estimates imply $\omega_{int}/\nu\approx 8.6$,  
which corresponds to 9 fast oscillations per one period of the slow beatings observed 
in Fig.~\ref{oscill}a.

\section{Interaction of a soliton with a defect.}
\label{sec_defect}

\begin{figure}[h]
\includegraphics[width=8.5cm,clip]{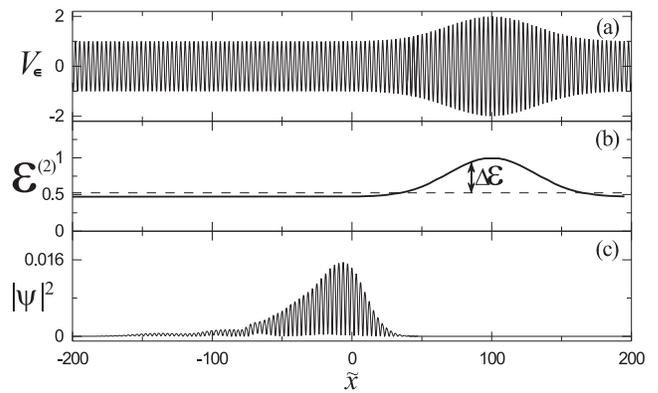}
\caption{(a) The lattice defect given by Eq.~(\ref{def1}) with $\epsilon=0.05$ and $\Delta \tilde{x}=100$. (b) The upper edge of the lowest gap (solid line) and the energy of the soliton with the initial velocity $v=0.1$ (dashed line). Intersections of these lines represent the turning points  (c) Profile of the reflected soliton   at 
time $t=150$. The initial soliton profile is the same as in Fig.~\ref{fig1}b.}
\label{fig5}
\end{figure}

The phenomenon of the  Bragg reflection of a soliton can be directly observed  in a system where 
the optical lattice has local increase of the depth, as it is shown in Figs.~\ref{fig5}a  
\begin{eqnarray}
\label{def1}
	V_\epsilon(\tilde{x})=\left[1+e^{-\epsilon^{5/2} (\tilde{x}-\Delta \tilde{x})^2}\right] 
\cos(2\tilde{x}). 
\end{eqnarray}
Local deformations of an optical lattice will be referred to as lattice defects. 
In this case, as before, we also start with a numerically obtained bright soliton, but adding now some initial 
velocity $v$ with respect to the stationary background which is introduced by means of the phase factor 
$ \psi(\tilde{x},0)e^{iv\tilde{x}/2}$. We are interested in the scattering process at different initial 
velocities, which are assumed to be  small enough: $v\ll 1$.  

In this case solitons are reflected from the defect (see a snapshot Fig.\ref{fig5}c). This is the effect of the  Bragg reflection which is explained in Fig.~\ref{fig5}b. Indeed, due to increase of the lattice depth, the gap is also increased locally. That is why the soliton, which initially was moving with a relatively small group velocity $v$, at some point reaches the (curved) band edge (it is given by the intersection of a dashed and solid lines in Fig.~\ref{fig5}b) which prevents further propagation of the carrier wave.
Thus, the defect described by the Eq.~(\ref{def1}) can be classified as repulsive.

\begin{figure}[h]
\includegraphics[width=8.5cm,clip]{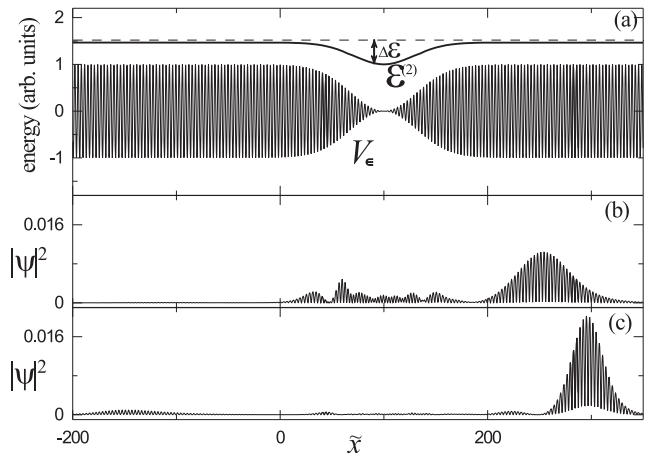}
\caption{(a) Periodic potential with the defect given by Eq.~(\ref{defect2}) and corresponding upper edge ${\cal E}^{(2)}$ of the first band gap (here the dashed line is initial soliton energy). Parameters of the defect are $\epsilon=0.05$ and $\Delta \tilde{x}=100$. Profiles of the soliton with different initial 
velocities $v=0.1$ (b), and  $v=0.3$ (c) at times $t=300$ and $t=190$,  respectively. The 
initial soliton profile is the same as in Fig.~\ref{fig5}.}
\label{def}
\end{figure}

Let us consider now an opposite situation, when the depth of the periodic potential is locally decreased 
(see Fig.~\ref{def}a)
\begin{eqnarray}
\label{def2}
V_{\epsilon}(\tilde{x})=\left[1-e^{-\epsilon^{5/2} (\tilde{x}-\Delta \tilde{x})^2}\right] \cos(2\tilde{x}),
 \label{defect2}	
\end{eqnarray}
what leads to local narrowing of the forbidden gap.  

By using the arguments based on the band gap structure outlined below one may argue
that the modulation described by the Eq.~(\ref{def2}) acts as an attractive impurity. Indeed, 
in the region of the defect there exists a forbidden band shrinking. Assuming as before, that the soliton 
energy ${\cal E}$ is constant, one concludes that in the region of the defect the energy shift 
 $\Delta {\cal E}$ (see Fig.~\ref{fig_add}) toward the allowed zone depends on the coordinate 
$\Delta {\cal E}=\Delta {\cal E}(\tilde{x})$: it increases as the soliton approaches the center of the 
defect and then decreases as coordinate increases. As far as larger 
$\Delta {\cal E}(\tilde{x})$ correspond to larger velocities of the carrier wave, the 
soliton is accelerating when it approaches the defect and decelerating when soliton moves outward the defect.  

Since the lattice modulation (\ref{def2}) acts as attractive a number of atoms could be captured by such a defect, in the case 
when an  initial kinetic energy of the condensate is small enough. This is exactly what we observe in numerical 
simulations shown in Fig.~\ref{def}b. The higher velocity matter waves pass through the defect without substantial 
changes   (see Fig.~\ref{def}b).

\section{Conclusion}
 
In conclusion, we have shown that in smoothly modulated optical lattices that can be created by using 
quasi-monochromatic laser beams, one can effectively manage matter solitons as they accelerate, 
deccelerate, oscillate or undergo the reflection depending on the type of the modulation introduced.  
Since the above processes are controlled by the periodic structure, not only 
dynamical, but also other properties of matter waves such as the energy, the effective mass and width of the soliton can be manipulated.
Although in a particular dynamical process a matter wave can loose its solitonic 
properties, the effective mass approach provides a qualitative explanation of the main features of 
the soliton dynamics, if the wave packet possesses a substantially larger extension than the lattice 
period and lattice modulations are smooth enough.

\section*{Acknowledgments}


Work of V.A.B. has been supported by the FCT
fellowship SFRH/BPD/5632/2001. V.V.K. acknowledges support from the European grant, COSYC n.o. HPRN-CT-2000-00158. 
V.K. acknowledge support from COST P11 Action. Cooperative work was supported by the bilateral agreement 
GRICES/Czech Academy of Sciences.

\end{document}